\documentclass[aps, unsortedaddress, superscriptaddress]{revtex4-1}
\pdfoutput=1
\usepackage[english]{babel}
\usepackage{geometry,amsmath,amssymb,graphicx,textcomp,latexsym}
\newcommand{\tmem}[1]{{\em #1\/}}

\newcommand{\tmnote}[1]{\thanks{#1}}
\newcommand{\tmop}[1]{\ensuremath{\operatorname{#1}}}
\newcommand{\tmstrong}[1]{\textbf{#1}}
\newcommand{\tmtextit}[1]{{\itshape{#1}}}
\newcommand{\tmverbatim}[1]{{\ttfamily{#1}}}
\newenvironment{itemizeminus}{\begin{itemize} }{\end{itemize}}
\begin{document}

\title{Microscopic charged fluctuators as a limit to the coherence of
disordered superconductor devices}

\author{H{\'e}l{\`e}ne le Sueur}
\altaffiliation[now at ]
{Quantronics \ group, \ SPEC, \ CEA, \ CNRS (UMR 3680), \
Universit{\'e} \ Paris-Saclay, \ CEA \ Saclay \ 91191 \ Gif-sur-Yvette \
Cedex, France}
\affiliation{CSNSM, Univ. Paris-Sud, CNRS/IN2P3, Universit{\'e} Paris-Saclay,
91405 Orsay, France}

\author{Artis Svilans}
\altaffiliation[now at ]{Division of Solid State Physics and NanoLund, Lund University,
Box 118,S-221 00 Lund, Sweden}
\affiliation{Quantronics \ group, \ SPEC, \ CEA, \ CNRS (UMR 3680), \
Universit{\'e} \ Paris-Saclay, \ CEA \ Saclay \ 91191 \ Gif-sur-Yvette \
Cedex, France}
\author{Nicolas Bourlet}
\affiliation{Quantronics \ group, \ SPEC, \ CEA, \ CNRS (UMR 3680), \
Universit{\'e} \ Paris-Saclay, \ CEA \ Saclay \ 91191 \ Gif-sur-Yvette \
Cedex, France}
\author{Anil Murani}
\affiliation{Quantronics \ group, \ SPEC, \ CEA, \ CNRS (UMR 3680), \
Universit{\'e} \ Paris-Saclay, \ CEA \ Saclay \ 91191 \ Gif-sur-Yvette \
Cedex, France}
\author{Laurent Berg{\'e}}
\affiliation{CSNSM, Univ. Paris-Sud, CNRS/IN2P3, Universit{\'e} Paris-Saclay,
91405 Orsay, France}
\author{Louis Dumoulin}
\affiliation{CSNSM, Univ. Paris-Sud, CNRS/IN2P3, Universit{\'e} Paris-Saclay,
91405 Orsay, France}

\author{Philippe Joyez}
\tmnote{Corresponding author, Email: \tmverbatim{philippe.joyez@cea.fr}}
\affiliation{Quantronics \ group, \ SPEC, \ CEA, \ CNRS (UMR 3680), \
Universit{\'e} \ Paris-Saclay, \ CEA \ Saclay \ 91191 \ Gif-sur-Yvette \
Cedex, France}

\begin{abstract}
  By performing experiments with thin-film resonators of NbSi, we elucidate a
  decoherence mechanism at work in disordered superconductors. This
  decoherence is caused by charged Two Level Systems (TLS) which couple to the
  conduction electrons in the BCS ground state; it does not involve any
  out-of-equilibrium quasiparticles, vortices, etc. Standard theories of
  mesoscopic disordered conductors enable making predictions regarding this
  mechanism, notably that decoherence should increase as the superconductor
  cross section decreases. Given the omnipresence of charged TLS in
  solid-state systems, this decoherence mechanism affects, to some degree, all
  experiments involving disordered superconductors. In particular, we show it
  easily explains \ the poor coherence observed in quantum phase slip
  experiments and may contribute to lowering the quality factors in some
  disordered superconductor resonators.
\end{abstract}

{\maketitle}

\paragraph*{Introduction.} In highly disordered superconductor films, the
kinetic inductance of carriers can exceed the geometrical inductance by orders
of magnitude. These materials have been proposed for fabricating new, purely
dispersive, high-impedance electronic devices such as tunable (super)inductors
{\cite{adamyan_tunable_2016,annunziata_tunable_2010-1}}, slow-wave
transmission lines {\cite{adamyan_kinetic_2015}}, photon detectors
{\cite{diener_design_2012,leduc_titanium_2010}} and coherent Quantum Phase
Slip Junctions (QPSJ){\cite{mooij_superconducting_2006-1}}, the latter being
the dual component of the celebrated Josephson Junction (JJ). The QPSJ
proposal has drawn particular interest, as it enables designing new
superconducting quantum circuits
{\cite{hriscu_model_2011-1,hriscu_coulomb_2011-1,kerman_metastable_2010,hriscu_quantum_2013-1}}
that operate in the previously inaccessible high impedance domain, and which
could find applications in quantum technologies. In the recent years, circuits
embedding QPSJ of different materials have been tested
{\cite{peltonen_hybrid_2018,peltonen_coherent_2016,arutyunov_junctionless_2017,peltonen_coherent_2013,astafiev_coherent_2012,graaf_charge_2018}},
but they all displayed low coherence times compared to those routinely
achieved in JJ-based circuits. In the present work, we bring to light a new
decoherence mechanism in highly disordered superconductor, due to the charged
TLS omnipresent at interfaces and in insulators in solid state systems. This
mechanism easily explains the poor coherence observed in QPSJ devices. It also
likely contributes to lower-than-expected quality factors reported in
resonators made with disordered superconductors
{\cite{astafiev_coherent_2012,peltonen_coherent_2016,samkharadze_high-kinetic-inductance_2016,maleeva_circuit_2018-1,zhang_microresonators_2018}},
even though this mechanism is non-dissipative. This mechanism should affect
all the proposals mentioned above, but also experiments probing the
Superconductor to Insulator quantum phase Transition (SIT)
{\cite{gantmakher_superconductorinsulator_2010-1,lin_superconductivity_2015}},
or the Berezinskii--Kosterlitz--Thouless
(BKT){\cite{konig_berezinskii-kosterlitz-thouless_2015,schneider_excess_2014,zhao_evidence_2013}}
phase transition.

Thin-films of various disordered superconductors have been used to implement
high kinetic inductance circuits. Provided they are not too close to the SIT
(occurring at a normal state sheet resistance $R_{N \Box}$ of order $h / 4 e^2
\simeq 6.5 ~ \text{k} \Omega$), these materials can still be qualitatively
described by the BCS theory. In this framework, well below the critical
temperature, the low-frequency linear response of a diffusive superconductor
is described by a kinetic inductance $L_K$ proportional to its normal state
resistance $R_N$ {\cite{fominov_surface_2011}},
\begin{equation}
  L_K = \frac{R_N}{\pi \omega_{\tmop{gap}}} \label{Lkin},
\end{equation}
where $\omega_{\tmop{gap}} = \Delta / \hbar$, with $\Delta$ the
superconducting gap. With thin films having $R_{N \Box} \sim 1 \text{k}
\Omega$, and given the common superconducting gaps are in the range 0.2-2.0
meV, one can then achieve sheet kinetic inductance $L_{K \Box}$ of the order
of 1 nH. Thus, these materials should enable highly inductive components that
operate up to frequencies of the order of $\omega_{\tmop{gap}}$ ({\tmem{i.e.}}
tens to hundreds of GHz) which are otherwise unfeasible.

\paragraph*{Experiments.} In this work we have used $\tmop{Nb}_x \tmop{Si}_{1
- x}$ alloy, with $x = 0.18$, and a film thickness $t = 15$ nm, deposited on
an intrinsic Si substrate. The alloy was co-evaporated as described in Ref.
{\cite{crauste_tunable_2011-1}}. In the normal state, this film has a sheet
resistance $R_{N \Box}$ of about 600 $\Omega$, weakly dependent on
temperature. The RF properties of the $\tmop{Nb}_{} \tmop{Si}$ layer were
first characterized by fabricating a half-wavelength coplanar waveguide
resonator with a 10 {\textmu}m-wide, 700 {\textmu}m-long central conductor and
measuring it at 17~mK, well below the critical temperature of $0.85$~K. From
the fundamental resonance (at 6.687~GHz), we extracted the sheet inductance
$L_{K \Box} = 0.83 ~ \tmop{nH}$. The measured quality factor $Q_{\tmop{meas}}
= 1.6 \times 10^4$ of the resonance was much lower than the designed external
quality factor $Q_{\tmop{ext}} = 1.6 \times 10^5$, set by the capacitive
coupling to the input and output $50 \Omega$ lines. Lower-than-expected
quality factors have been reported in several resonators made of disordered
superconductors
{\cite{astafiev_coherent_2012,peltonen_coherent_2016,samkharadze_high-kinetic-inductance_2016,maleeva_circuit_2018-1}};
they are usually attributed to unspecified ``internal losses'', although
dissipative mechanisms are unexpected at these temperatures in BCS
superconductors. Finally, the non-linear response to the amplitude of the
probe signal gave access to the superconducting coherence length $\xi \simeq
40 ~ \tmop{nm}$ of the material (see Ref. {\cite{bourlet_preparation_2018}}
for details).

Using the above value for $L_{K\Box}$, we designed lumped-element $\tmop{LC}$
resonators in which the inductors were narrow wires 100-180~nm in width, and
the capacitors were rectangular pads at both ends of the inductor (see Fig.
1a). The dimensions were chosen with the aid of numerical simulations in order
to produce well-separated resonance frequencies between 6 and 8 GHz (see Table
1). The resonators were all capacitively coupled to the same input and output
50 $\Omega$ transmission lines, weakly enough to all have quality factors
larger than $10^4$ in absence of internal losses. The design was transferred
into the NbSi layer using e-beam lithography and dry etching in a $\tmop{CF}_4
- \tmop{Ar}$ mixture with the negative ma-N resist acting as a mask. After
removal of the mask, the sample was wire-bonded on a printed-circuit board and
cooled down in a dilution refrigerator. As shown in Fig. 1a, the wiring of the
sample incorporated a bias tee and a DC voltage source letting us apply an
electric field on the resonators.

\begin{figure}[h]
  \resizebox{15cm}{!}{\includegraphics{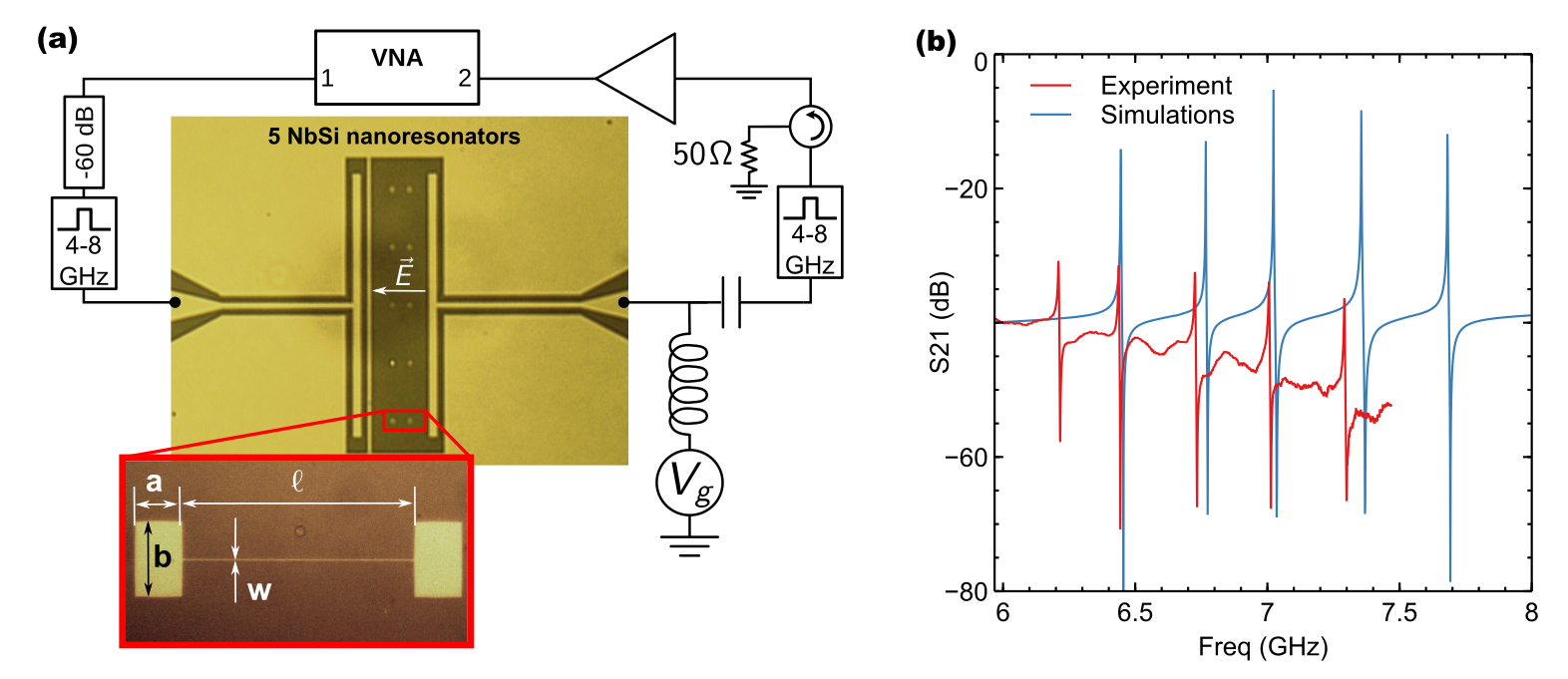}}
  \caption{{\tmstrong{(a)}} Simplified schematic of the setup. Five different
  $\tmop{NbSi}$ nanoresonators are measured in transmission, using a Vector
  Network Analyzer. A bias tee and a voltage source $V_g$ enable applying an
  electric field along the resonators. {\tmstrong{(b)}} Simulated (assuming
  zero internal losses) and measured transmission $S_{21}$ of the sample. The
  resonance peaks have a Fano resonance shape due to the stray direct coupling
  of the input and output transmission lines in addition to the coupling
  through the resonator. The experimental curve is offset vertically to
  account for the total attenuation and amplification of the setup at 6 GHz.
  Experimental resonances reach much lower peak transmissions than the
  simulations, usually indicating that internal losses dominate. }
\end{figure}

\begin{table}[h]
  \begin{tabular}{l}
    
  \end{tabular}\begin{tabular}{|c|c|c|c|c|c|c|c|c|}
    \hline
    Resonator & $w (\tmop{nm})$ & $\ell$ ({\textmu}m) & $a ({\textmu}m)$ & $b
    ({\textmu}m)$ & $f_{\tmop{des}}$ (GHz) & $f_{\tmop{meas}}$ (GHz) &
    $Q_{\tmop{ext}} $ & $Q_{\tmop{meas}}$\\
    \hline
    1 & 100 & 50 & 10 & 10 & 6.44 & 6.21 & 17100 & 1890\\
    \hline
    2 & 120 & 50 & 12 & 10 & 6.76 & 6.44 & 20900 & 2280\\
    \hline
    3 & 140 & 50 & 14 & 10 & 7.02 & 6.73 & 14000 & 2020\\
    \hline
    4 & 160 & 50 & 15 & 10 & 7.35 & 7.00 & 12900 & 2140\\
    \hline
    5 & 180 & 50 & 16 & 10 & 7.68 & 7.29 & 16000 & 2650\\
    \hline
  \end{tabular}
  \caption{Design parameters of the five resonators and results of their
  experimental characterization (see Fig. 1a for resonator dimensions).
  $f_{\tmop{des}}$ is the designed resonance frequency and $Q_{\tmop{ext}}$ is
  the designed external quality factor (i.e. simulated total quality factor,
  with no internal losses). $f_{\tmop{meas}}$ and $Q_{\tmop{meas}}$ are the
  measured resonance frequency and total quality factor, obtained by fitting
  $S_{21}$.}
\end{table}

In Fig. 1b we show the simulated and measured transmission $S_{21}$ through
the five resonators. The measurements were done at 30 mK, with sufficiently
low power for the resonators to be well within their linear response regime.
We observe that the resonance frequencies are well-spaced, as designed, with,
however, a systematic shift towards lower frequencies which we attribute to a
slightly non-nominal fabrication process. We also observe that the measured
resonance peaks are markedly lower than the simulated values. Correspondingly,
the measured quality factors $Q_{\tmop{meas}}$, obtained by fitting the
$S_{21}$ data as a function of frequency, are much lower than the external
quality factors $Q_{\tmop{ext}}$ predicted by simulations (see Table 1). We
further observe that $Q_{\tmop{meas}}$ is about an order of magnitude lower
than what had been determined in the large half-wavelength resonator. Since
the small and large resonators were made out of the same NbSi layer and
following nominally the same process, the large decrease in $Q_{\tmop{meas}}$
suggests the quality factor in NbSi resonators depends on the lateral
dimension, and is higher in wider structures.

In Fig. 2a we show the variations of the phase of $S_{21}$ at a fixed
frequency $f = 6.2086 \tmop{GHz}$, close to the maximum transmission of
resonator \#1, while the gate voltage is repeatedly swept from 0 to 110 mV
over 3.5 s. In each sweep, we observe several abrupt changes in the phase. The
gate voltage at which these jumps occur varies from one sweep to another,
showing slow drifts and telegraphic signal-like jumps. Furthermore, such
measurements repeated at different times displayed long-term variability
typical of $1 / f$-like noise. Figure 2b shows that the observed flicker noise
corresponds to fluctuations in the {\tmem{resonance frequency}}. Similar
features were observed in the other 4 nanoresonators. Due to the frequency
fluctuations, the phase of the (electromagnetic) quantum state of the
resonator becomes unknown after some time (the {\tmem{dephasing}} time) and
the state can no longer be manipulated deterministically --a phenomenon known
as {\tmem{dephasing}} or \tmtextit{decoherence}.

\begin{figure}[h]
  \resizebox{15cm}{!}{\includegraphics{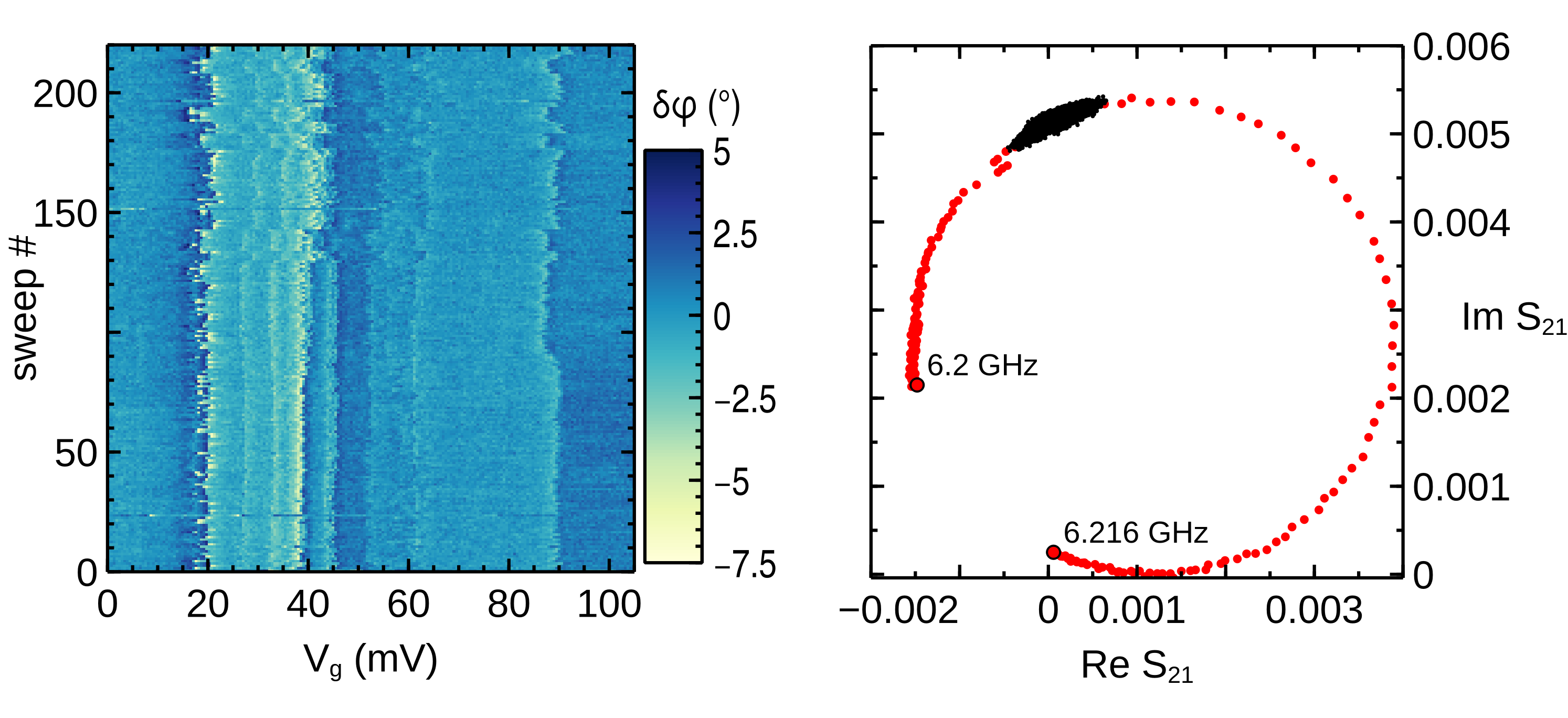}}
  \caption{{\tmstrong{(a)}} Variations of the phase of $S_{21}$ at a fixed
  frequency $f = 6.2086 \tmop{GHz}$, close to the maximum transmission of
  resonator \#1, while the gate voltage is repeatedly swept from 0 to 110 mV
  over 3.5 s {\tmstrong{(b)}} $S_{21}$ in the complex plane for resonator \#1.
  Red dots were obtained by varying the frequency across the resonance. The
  cloud of black dots corresponds to all the data points in {\tmstrong{(a)}}.
  It is aligned along the resonance circle, showing that the jumps observed
  are really resonance frequency jitter.}
\end{figure}

\paragraph*{Interpretation.}It is tempting to attribute the fluctuations in
the resonance frequency of our nanoresonators to the mechanisms already known
to occur in superconducting resonators. Indeed, it was understood in the
recent years that microscopic charge systems present in the dielectric
material surrounding superconducting resonators cause frequency jitter and
losses in these resonators
{\cite{gao_physics_2008,pappas_two_2011,graaf_suppression_2018,quintana_observation_2017}},
on the grounds of the so-called Generalized Tunneling Model (GTM)
{\cite{faoro_interacting_2015}}. In this model, the losses are due to TLSs
resonantly coupled to the resonator via the AC electric field, each of these
TLSs also interacting strongly with other, non-resonant, thermally activated
TLSs. Overall, these couplings result in fluctuations of the dielectric
constant at the resonant frequency, i.e. fluctuations in the effective
capacitance of the resonator. While this TLS-induced dielectric loss mechanism
is certainly present in our experiments, one does not expect it to yield
observable individual TLS switching events. This is because in our geometry,
any TLS occupies an extremely small volume fraction of the resonator mode, and
it can only modify the capacitance accordingly. Coupling to spin impurities
(magnetic TLSs) may also be a source of fluctuations in superconducting
resonators
{\cite{de_graaf_direct_2017,samkharadze_high-kinetic-inductance_2016}}, but
strong resonant magnetic coupling with individual localized spin impurities is
similarly ruled out.

\paragraph*{A new dephasing mechanism.} Here, we propose an alternative
explanation for the observed fluctuations. We argue it is the direct
interaction of the TLSs with the conduction electrons which causes flicker
noise of the kinetic inductance. In the GTM {\cite{faoro_interacting_2015}},
the corresponding interaction Hamiltonian for a single TLS reads
\begin{equation}
  H_{\text{TLS-el}} = \sigma_z  \sum_{kk' \eta} V_{kk'} c_{k \eta}^{\dagger}
  c_{k' \eta} \label{Hint}
\end{equation}
where $V_{kk'}$ describes the scattering potential, $c_{k \eta}^{\dagger}
(c_{k \eta})$ creates (annihilates) a fermion of wave vector $k$ and spin
$\eta$, and $\sigma_z$ is the Pauli matrix describing the TLS. However, in
Ref. {\cite{faoro_interacting_2015}} this term is only considered for its
relaxation effect on the TLS, neglecting the corresponding back-action on the
conduction electrons. We show below this is legitimate in conventional
superconducting resonators in which the kinetic inductance is a negligible
part of the total inductance (it happens to be the case for all experiments
where TLS losses were carefully analyzed
{\cite{gao_physics_2008,pappas_two_2011,graaf_suppression_2018,quintana_observation_2017}}),
but not in highly disordered superconductors.

For simplicity, we first consider the case where the length $\ell$ of the
inductor wire in our resonators is shorter than the electronic coherence
length $L_{\varphi}$ in NbSi and use the Landauer-B{\"u}ttiker framework
{\cite{heikkila_physics_2013}} to describe transport. The normal-state
conductance of the wire is given by the Landauer formula,
\[ G_N = 2 G_0 \sum_n T_n \]
where $G_0 = e^2 / h$, and the $T_n$ are the transmission probabilities of the
channels, i.e. the square modulus of the transmission eigenvalues. For
disordered materials like our wires, the transmission results from
multiple-path interferences though the wire, analogous to a speckle pattern in
optics. In other words, the $T_n$ depend in an intricate manner on the
specific disorder realization which is unknown. Theory
{\cite{heikkila_physics_2013}} nevertheless provides statistical predictions
regarding $G_N$, namely that it has
\begin{itemizeminus}
  \item an expectation value related to the macroscopic sheet resistance of
  the material and wire dimensions
  \begin{equation}
    \langle G_N \rangle = \frac{w}{\ell R_{N\Box}} \label{GN}
  \end{equation}
  \item a standard deviation $\sigma_G$ according to Universal Conductance
  Fluctuations (UCF), \tmtextit{i.e.} $\sigma_G \sim G_0$ in the ``metallic
  case'' ($G_N > G_0$), and slowly decreasing approximately as $\sigma_G \sim
  G_0 \times \sqrt{G_N / G_0}$ in the crossover to the Anderson insulator
  regime ($G_N < G_0$) {\cite{qiao_universal_2010}}, independently of the wire
  material or dimensions.
\end{itemizeminus}
When a TLS changes state, it locally modifies the electronic scattering
potential in its vicinity, imparting new phaseshifts to the electronic
trajectories. For the whole wire, this results in a change of the global
electronic speckle pattern, i.e. a modification of the channels, with a change
of the conductance $G_N \rightarrow G_N + \delta G$. If the TLS is located far
away from the wire, the change in the potential is vanishingly small and
$\delta G = 0$. In the opposite limit of ``strong interaction'' where the
switching of the TLS radically changes the speckle pattern, $\delta G$ has a
random value with a standard deviation constrained by the UCF. Hence, for any
given TLS one expects $0 \leqslant | \delta G | \lesssim G_0$ depending on the
type of TLS and its distance to the wire. In other words, the random
conductance jumps of the individual TLSs follow a distribution, itself derived
from the distribution of the coupling strengths (hereafter denoted
$\mathcal{D}_{\tmop{cs}}$). Another important characteristic of a TLS is its
switching time; for the TLS ensemble, switching times are random, following a
distribution $\mathcal{D}_{\tmop{st}}$.

In the superconducting state, $\delta G$ causes a change of the kinetic
inductance of the wire (Eq. (\ref{Lkin}) with $R_N = G_N^{- 1}$). This has an
effect on the resonance frequency $f = 1 / 2 \pi \sqrt{(L_K + L_{\tmop{geom}})
C}$ with $C$ the capacitance of the resonator and $L_{\tmop{geom}}$ the
geometrical inductance. To first order, a TLS Switching Event (SE) induces a
relative change in the resonance frequency
\begin{equation}
  \frac{\delta f}{f} = - \frac{\alpha}{2}  \frac{\delta L_K}{L_K} =
  \frac{\alpha}{2}  \frac{\delta G}{G_N} \label{deltaf},
\end{equation}
where $\alpha = L_K / (L_K + L_{\tmop{geom}})$ is the participation ratio of
$L_K$ in the total inductance. While $\alpha \simeq 1$ in our disordered
resonator, it is vanishingly small in ``conventional'' superconducting
resonators which are then not dephased by TLS through this mechanism (the
dielectric losses of the GTM then being dominant).

Successive single TLS SEs produce a random-walk-like evolution for the
conductance (bounded by UCF) and of the resonance frequency. Measured over
many SEs, this leads to an increased resonance width $\Delta f_{\tmop{FWHM}}$
accompanied with a reduction of the peak transmitted power because the
resonator never stays long in optimum transmission condition. Assuming the TLS
fluctuations are the main cause of the resonance width $\Delta
f_{\tmop{FWHM}}$, the corresponding quality factor $Q_{\tmop{TLS}} = f /
\Delta f_{\tmop{FWHM}}$ is given by
\begin{equation}
  Q_{\tmop{TLS}}^{- 1} \sim \frac{2 \langle \delta f \rangle_{\tmop{rms}}}{f}
  = \alpha \frac{\langle \delta G \rangle_{\tmop{rms}}}{G_N} \label{Q}
\end{equation}
where $\langle \ldots \rangle_{\tmop{rms}}$ denotes the rms-averaged quantity
over all single TLS jumps during the measurement. $\langle \delta G
\rangle_{\tmop{rms}}$ can be obtained from the distributions
$\mathcal{D}_{\tmop{cs}}$ and $\mathcal{D}_{\tmop{st}}$. The UCF upper bound
$\langle \delta G \rangle_{\tmop{rms}} = \sigma_G \sim G_0$ is reached when
the speckle pattern is sufficiently reorganized during the measurement,
requiring sufficiently strongly coupled and fast TLSs. From Eqs.
(\ref{deltaf}) and (\ref{Q}), one expects a larger effect of TLSs in systems
with a smaller conductance, \tmtextit{i.e.} a smaller number $N$ of conduction
channels, as $N \propto G_N \propto w \times t$.

\paragraph*{Microscopic and macroscopic dephasing.} The above discussion
brings to light a TLS-Induced Dephasing Mechanism (TLSIDM) in disordered
superconductor resonators, i.e. the dephasing of collective, macroscopic,
electromagnetic degrees of freedom in the circuit. These results were obtained
considering that electrons in the inductors are fully phase-coherent. However,
this assumption is unrealistic because our inductors are longer $\left( \ell =
50 \text{{\textmu}m} \right)$ than the largest published values for
$L_{\varphi}$ in metals. Moreover, TLS SEs themselves introduce random
electronic phase shifts, contributing to shortening $L_{\varphi}$. In the
following we take into account such (microscopic) dephasing of individual
electrons in our analysis, and work out how it modifies the (macroscopic)
TLSIDM.

To this effect, we first examine electronic dephasing only. On average, SEs
dephase the electrons after a time $\tau_{\tmop{TLS}}$ stemming from
$\mathcal{D}_{\tmop{cs}}$ and $\mathcal{D}_{\tmop{st}}$ (which already
determine $\langle \delta G \rangle_{\tmop{rms}}$). Assuming TLSs are the main
source of electron dephasing, one then has $L_{\varphi} = \sqrt{D
\tau_{\tmop{TLS}}}$ where $D$ is the diffusion constant. Yet, the value of
$L_{\varphi}$ in the experiment is not known, partly because the standard weak
localization determination cannot be performed in the superconducting state.
How does this dephasing of individual electronic states affect the collective
superconducting state (i.e. the BCS state)? As long as $\tau_{\tmop{TLS}}$ is
large compared to the timescale of the pairing interaction $\hbar / \Delta$
(i.e. $L_{\varphi} \gg \xi$), the superconducting order adapts to such
fluctuations and $| \Delta |$ remains essentially unchanged, just as if the
disorder were static (Anderson's theorem). The phase of the order parameter,
however, is a macroscopic electromagnetic degree of freedom which is affected
by TLS-induced dephasing proportionally to $\alpha$, as in Eq. (\ref{deltaf}).
In the opposite limit $\tau_{\tmop{TLS}} < \hbar / \Delta$, fast electron
dephasing simply inhibits the formation of the superconducting state. In the
crossover, one expects $| \Delta |$ to be reduced and the density of states to
be modified, which could be related to anomalous properties reported in some
disordered superconductor resonators {\cite{driessen_strongly_2012-1}}. In the
following, for simplicity, we assume $L_{\varphi} \gg \xi$.

How does the finite coherence length of electrons modify the TLSIDM? In a
quasi-1D incoherent inductor wire (\tmtextit{i.e.} with $w, t \ll L_{\varphi}
< \ell$), an individual TLS SE modifies the normal state conductance of the
wire only over an $L_{\varphi}$-long segment, yielding a $\delta G$ typically
smaller by a factor $(L_{\varphi} / \ell)^2$ than in the coherent case
(\tmtextit{i.e.} $0 \leqslant | \delta G | \lesssim (L_{\varphi} / \ell)^2
G_0$). The resistance changes corresponding to SEs in different segments add
in a random walk fashion, so that, averaged over many TLS configurations,
$\langle \delta G \rangle_{\tmop{rms}}$ is smaller than in a coherent wire by
a factor $(L_{\varphi} / \ell)^{3 / 2}$, just like $\sigma_G$ is reduced in
incoherent mesoscopic conductors {\cite{heikkila_physics_2013}}. Assuming
strongly coupled TLSs, one then gets the upper limit for $\langle \delta G
\rangle_{\tmop{rms}}$
\begin{equation}
  \langle \delta G \rangle_{\tmop{rms}} \sim \left( \frac{L_{\varphi}}{\ell}
  \right)^{3 / 2} G_0 \label{strongnoise}
\end{equation}
(recalling however that $L_{\varphi}$ and $\langle \delta G
\rangle_{\tmop{rms}}$ are not fully independent). Then, using Eqs. (\ref{Q})
and (\ref{GN}), this yields the strong-coupling prediction
\begin{equation}
  Q_{\tmop{TLS}}^{- 1} \simeq \alpha \left( \frac{L_{\varphi}}{\ell}
  \right)^{3 / 2} \frac{\ell}{w} G_0 R_{\Box} \label{strongcouplingQ},
\end{equation}
setting an absolute lower bound for the quality factor due to TLSIDM.

\paragraph*{Comparison with our experiments.} The above analysis is in
qualitative agreement with our observations. In the first place, the mechanism
we propose can explain the observed telegraphic-like frequency jitter, the
resulting dephasing and the lower-than-expected quality factors. In the second
place, the prediction of larger relative frequency fluctuations in systems
with fewer conduction channels explains why the nanoscale resonators appear
more lossy than the larger resonator, although they are made of the same
material, with (nominally) the same density of volume and surface defects. The
data from our 5 nanoresonators is also qualitatively compatible with the
proportionality between the quality factor and the nanowire width (Eqs.
(\ref{Q}) \& (\ref{GN})), but there are too few samples and a too narrow range
of widths to regard this as a solid demonstration. \

Let us now try to be more quantitative. The largest jumps in Fig. 2 are of
the order of 10{\textdegree}, which, given the $S_{21}$ resonance circle,
corresponds to $\delta f \simeq 0.6 ~ \tmop{MHz}$, and, according to Eq.
(\ref{deltaf}), a change in the normal state conductance $| \delta G | \simeq
1.4 \times 10^{- 5} G_0$. Since we have selected the largest jump observed, we
boldly assume it corresponds to a strongly coupled TLS inducing a change of
conductance of the order of the disorder-averaged rms value $\sim ~
(L_{\varphi} / \ell)^2 G_0$, which then yields an estimate of $L_{\varphi}
\sim 0.2 \text{{\textmu}m}$. Similarly, if we assume a strongly coupled TLS
ensemble, then, matching the measured quality factor $Q_{\tmop{meas}} \sim 2
\times 10^3$ with $Q_{\tmop{TLS}}$ in Eq. (\ref{strongcouplingQ}) yields
$L_{\varphi} \sim 0.6 \text{{\textmu}m}$, which is of the same order than the
above crude estimate. This assumption of strongly coupled TLSs yields a lower
bound for $L_{\varphi}$. It could be that there are no strongly coupled TLS in
our resonators ($| \delta G | \ll (L_{\varphi} / \ell)^2 G_0$ for all
individual TLSs), in which case the observations would be consistent with a
larger $L_{\varphi}$.

\paragraph*{TLS types.}We now discuss the type of TLS that cause such
decoherence. As NbSi is an amorphous material, TLSs may be atoms jumping
between metastable positions in the bulk of the material. However, the
observed effect of a DC electric field proves that charged traps or dipolar
defects in the vicinity of the wire are at work. It is indeed well known that
such charged TLSs are omnipresent in insulators or at interfaces in
solid-state electronic devices where, besides the GTM dielectric noise already
mentioned {\cite{faoro_interacting_2015}}, they cause charge noise by coupling
to surface electronic states (within a Thomas-Fermi screening length
$\lambda_{\tmop{TF}}$). This charge noise is well documented in mesoscopic
circuits such as single electron transistors
{\cite{gustafsson_thermal_2013,zorin_background_1996}}, charge qubits
{\cite{vion_manipulating_2002,ithier_decoherence_2005}} or quantum point
contacts {\cite{halbritter_slow_2004}}, but also in MOS transistors
{\cite{fan_single-trap-induced_2014}}. The microelectronics industry has shown
that a careful choice of materials and process engineering can much reduce the
number of charged TLSs, but never completely suppress them. Note that besides
the TLSIDM dependence on the number of channels $N \propto w \times t$, for
charged TLSs one further expects the TLSIDM to become stronger when the
thickness $t$ is reduced to become comparable to $\lambda_{\tmop{TF}}$, as
charged TLSs then interact with a larger fraction of conduction electrons.

\paragraph*{Comparison with other experiments.} The TLSIDM we point out also
directly affects QPS experiments in disordered superconductors. The phase slip
energy of a uniform quasi-1D nanowire is predicted
{\cite{mooij_superconducting_2006-1,zaikin_quantum_1997}} to scale as
\begin{equation}
  E_S = \Delta \frac{G_N}{G_0} \left( \frac{\ell}{\xi} \right)^2 \exp \left( -
  A \frac{G_N}{G_0}  \frac{\ell}{\xi} \right) \label{Es}
\end{equation}
where $\Delta$ is the superconducting gap energy, $\xi$ the superconducting
coherence length, $\ell$ the nanowire length, and $A$ a numerical factor of
order 1. In order to have non vanishing $E_S$, one needs nanowires with a
small number of conduction channels. In Ref. {\cite{peltonen_coherent_2013}},
it was already noticed that the randomness in disorder realizations induces a
relatively large (static) dispersion in values of $G_N$ of order $G_0$,
resulting in an irreproducibility of $E_S$ among nominally identical samples.
It was also pointed that random offset charges (i.e. essentially frozen
charged TLS) were likely contributing to this disorder. Here, we simply extend
this reasoning, arguing that charged TLS in the vicinity of the wire also
induce {\tmem{dynamical}} fluctuations in $E_S$. These fluctuations of $E_S$
are dual to fluctuations in the Josephson energy of JJs due to TLS in the
tunnel barrier {\cite{simmonds_decoherence_2004}}, both resulting in
decoherence.

We now examine whether the TLSIDM we propose could explain a few published
experimental results. In Ref. {\cite{maleeva_circuit_2018-1}}, using the
values given for the granular Aluminum resonator shown in Fig. 3a, the
measured quality factor would be consistent with $Q_{\tmop{TLS}}^{}$ of
strongly coupled TLSs (Eq. (\ref{strongcouplingQ})) provided $L_{\varphi} \sim
0.5 \text{{\textmu}m}$. In Ref. {\cite{astafiev_coherent_2012}}, using the
values given for the $\tmop{InO}_x$ QPSJs and Eq. (\ref{Es}), one predicts
that a conductance noise $\langle \delta G \rangle_{\tmop{rms}} \sim 1.4
\times 10^{- 2} G_0$ due to charge fluctuators would yield a $\langle \delta
E_S \rangle_{\tmop{rms}} / E_S \sim$5\% fully explaining the observed Gaussian
spectroscopic linewidth. Strongly coupled TLSs (Eq. (\ref{strongnoise})) would
yield such $\langle \delta G \rangle_{\tmop{rms}}$ figure provided
$L_{\varphi} \simeq$15 nm, however, this seems exaggeratedly low for the
quoted $\xi = 30 \tmop{nm}$. Similar analysis carried out for the NbN QPSJs in
Ref. {\cite{peltonen_coherent_2013}}, the NbTiN resonators in Ref.
{\cite{samkharadze_high-kinetic-inductance_2016}}, or the granular Aluminum
resonator \#4 in Ref. {\cite{zhang_microresonators_2018}} all lead to
similarly low estimates of $L_{\varphi} \sim \xi$. As discussed in the case of
our resonator, all these experiments could also (more realistically)
correspond to weakly coupled TLS and larger $L_{\varphi}$. Thus, given a
suitable ensemble of charged TLS, the mechanism we propose could entirely
explain the observed decoherence in all these devices. On the other hand, in
some of these devices other mechanisms are clearly contributing to the overall
decoherence (in which case the TLSIDM must be weaker than we have just
estimated). For instance, in Ref.
{\cite{samkharadze_high-kinetic-inductance_2016}} spin impurities are shown to
play an important role; for the granular Al resonator of Ref.
{\cite{maleeva_circuit_2018-1}}, out-of-equilibrium quasiparticles have been
identified as a limiting factor {\cite{grunhaupt_loss_2018}}, and in Ref.
{\cite{zhang_microresonators_2018}} the power dependence of the quality factor
suggests that TLS dielectric losses of the GTM contribute to the resonance
width. Note however that, in the latter Ref.
{\cite{zhang_microresonators_2018}}, telegraphic fluctuations in the resonance
frequency are reported. This is similar to what we observe in our resonators
and, for the same reasons, we believe this is a manifestation of the TLSIDM.

\

\paragraph*{Discussion.} Let us address or clarify a few points regarding the
TLSIDM presented above.
\begin{itemizeminus}
  \item In resonators, internal energy losses (with an energy decay rate
  $\kappa_{\tmop{int}} > 0$) imply a lowering of the quality factor. The
  reciprocal connection is generally assumed to exist ({\tmem{i.e.}}
  lower-than-expected $Q \Rightarrow \kappa_{\tmop{int}} > 0$), but the TLSIDM
  is a counter-example disproving it. Indeed, this mechanism operates within
  the BCS ground state and it is thus strictly non-dissipative
  ($\kappa_{\tmop{int}} = 0$). If it were possible to measure fast enough
  between TLS switching events, one would observe at all times the narrow
  resonance limited by $Q_{\tmop{ext}}$, with the peak frequency jumping
  randomly. It is only after averaging over many TLS configurations
  ({\tmem{e.g.}} due to the measurement bandwidth) that one gets the reduced
  quality factor (Eq. (\ref{Q})), mimicking what internal losses would yield.
  As well, in Fig. 1b, the fact that the measured peak transmission is much
  lower than in the lossless simulations does not indicate energy being
  dissipated in the resonator, but power reflected to the input line when the
  TLSs shift the resonator out of resonance with the probe frequency. Given
  the jump dynamics involved in the disguise of the dissipationless mechanism
  as a lossy mechanism, in dubious cases, the relevant mechanism could be
  diagnosed by measuring with sufficient bandwidth the time-resolved
  transmitted power at a fixed frequency.
  
  \item The TLSIDM does not require any out-of-equilibrium quasiparticles,
  vortices, etc. which are usually invoked to explain decoherence in
  superconducting systems. The presence of such features would of course open
  additional decoherence channels.
  
  \item Although the TLSIDM we discuss is lossless, the GTM
  {\cite{faoro_interacting_2015}} shows that the TLS ensemble provides a
  dissipative bath able to absorb the resonator's microwave photons. This
  {\tmem{different}} GTM mechanism does yield internal energy losses (and
  decoherence) which are not taken into account in the present work, but which
  may contribute in some experiments. A distinctinctive feature in the GTM
  predictions is the non-monotonic dependence of losses on microwave power and
  temperature, due to the saturation of the resonant TLS. For the TLSIDM, TLSs
  couple to the wire independently of their energy splitting. Thus, for an
  ensemble of weakly coupled TLSs $(| \delta G | \ll (L_{\varphi} / \ell)^2
  G_0)$, we rather expect that the frequency fluctuations (respectively, the
  quality factor) due to this mechanism should decrease (resp., increase)
  monotonically when lowering the temperature, because less and less TLSs get
  thermally excited.
  
  \item It is well known that disordered superconductors develop spatial
  inhomogeneities in $| \Delta |$ {\cite{sacepe_disorder-induced_2008}} that
  eventually dominate the properties of the material close to the SIT
  {\cite{feigelman_microwave_2018}}. Since we only consider superconductors
  sufficiently far away from the SIT, we assume these inhomogeneities remain
  small and we have not taken them into account (e.g. in Eq. (\ref{Lkin})),
  but accounting for them would not qualitatively modify the TLSIDM we
  describe.
  
  \item Could the effect of charge noise be engineered-out for achieving
  highly coherent disordered superconductor circuits? After all, in JJ-based
  qubits this was achieved by operating the Cooper pair box at a ``sweet
  spot'' {\cite{vion_manipulating_2002,ithier_decoherence_2005}}, or by using
  the Transmon design
  {\cite{koch_charge-insensitive_2007,schreier_suppressing_2008}} in which
  charge sensitivity is suppressed exponentially. However, in these qubits,
  charge noise affects the charge degree of freedom in the system Hamiltonian,
  while in disordered superconductors charge noise modulates a parameter
  ($L_K$) of the Hamiltonian and, therefore, it cannot be mitigated in a
  similar way. Short of reducing TLS charge noise itself to an acceptable
  level, the only way to reduce the effect of TLSIDM is to reduce accordingly
  the kinetic inductance participation ratio $\alpha$, giving up at the same
  time the desired high kinetic inductance properties of the circuit. As an
  alternative, one may wonder if 1D topological superconducting channels,
  which are impervious to fluctuations in scattering, could form a basis for
  decoherence-free QPSJs or other high kinetic inductance devices.
\end{itemizeminus}
\paragraph*{Conclusions.} Because of the omnipresence of charged TLS, the
TLSIDM we describe is certainly at work in disordered superconductors devices.
We show it can easily explain the disappointing coherence of QPSJ, and that it
is likely contributing to the lower-than-expected quality factors of some
disordered superconductor resonators. Our results call for more in-depth
theoretical and experimental explorations of this TLSIDM, in order to
determine the expected distributions $\mathcal{D}_{\tmop{cs}}$ and
$\mathcal{D}_{\tmop{st}}$ for known TLSs, and obtain quantitative predictions
for the various noise spectra, coherence times, temperature dependence, etc.

Finally, beyond disordered superconductors devices, we also ponder that
fluctuators could have some impact on experiments investigating the SIT
{\cite{gantmakher_superconductorinsulator_2010-1,lin_superconductivity_2015}}
or BKT
{\cite{konig_berezinskii-kosterlitz-thouless_2015,schneider_excess_2014,zhao_evidence_2013}}
phase transitions in these materials. For instance, it is plausible that the
recently reported {\cite{kremen_imaging_2018}} telegraphic-like
reconfigurations in a disordered superconductor are due to the coupling to
charged TLS described above. As far as we know, all the theories for these
phase transitions assume a static disorder, and one may thus ask if and how
the TLS-induced dynamics would modify the characteristics of these
transitions. Then, the different densities and types of TLS in various
experiments could perhaps explain departures from the expected universality of
the SIT {\cite{lin_superconductivity_2015}}.

\

\paragraph*{Acknowledgments.} The authors are grateful to Genevieve Fleury,
Marc Westig and SPEC's Quantronics Group and Nanoelectronics Group members for
discussions, comments and support. This work was supported in part by ANR
grant ANR-15-CE30-0021-01, and by PALM and P2IO interLabex project NDS-NbSi. \

\end{document}